\documentclass[preprint]{aastex}
\usepackage{longtable}

\slugcomment{Accepted by the ApJS on April 11, 2016}

\shorttitle{THE AVERAGE IRON CHARGE STATE DISTRIBUTIONS}

\shortauthors{Song et al.}

\begin{document}
\title{A STATISTICAL STUDY OF THE AVERAGE IRON CHARGE STATE DISTRIBUTIONS INSIDE MAGNETIC CLOUDS FOR SOLAR CYCLE 23}
\author{H.Q. SONG\altaffilmark{1}, Z. ZHONG\altaffilmark{1}, Y. CHEN\altaffilmark{1}, J. ZHANG\altaffilmark{2}, X. CHENG\altaffilmark{3}, L. Zhao\altaffilmark{4}, Q. HU\altaffilmark{5}, AND G. LI\altaffilmark{5}}

\affil{1 Shandong Provincial Key Laboratory of Optical Astronomy
and Solar-Terrestrial Environment, and Institute of Space
Sciences, Shandong University, Weihai, Shandong 264209, China}
\email{hqsong@sdu.edu.cn}

\affil{2 School of Physics, Astronomy and Computational Sciences,
George Mason University, Fairfax, VA 22030, USA}

\affil{3 School of Astronomy and Space Science, Nanjing
University, Nanjing, Jiangsu 210093, China}

\affil{4 Department of Atmospheric, Oceanic, and Space Sciences,
University of Michigan, Ann Arbor, MI 48105, USA}

\affil{5 Department of Space Science and CSPAR, University of
Alabama in Huntsville, Huntsville, AL 35899, USA}


\begin{abstract}
Magnetic clouds (MCs) are the interplanetary counterpart of
coronal magnetic flux ropes. They can provide valuable information
to reveal the flux rope characteristics at their eruption stage in
the corona, which are unable to be explored in situ at present. In
this paper, we make a comprehensive survey of the average iron
charge state ($<$Q$>$Fe) distributions inside 96 MCs for solar
cycle 23 using \textit{ACE (Advanced Composition Explorer)} data.
As the $<$Q$>$Fe in the solar wind are typically around 9+ to 11+,
the Fe charge state is defined as high when the $<$Q$>$Fe is
larger than 12+, which implies the existence of a considerable
amount of Fe ions with high charge states (e.g., $\geq$16+). The
statistical results show that the $<$Q$>$Fe distributions of 92
($\sim$96\%) MCs can be classified into four groups with different
characteristics. In group A (11 MCs), the $<$Q$>$Fe shows a
bimodal distribution with both peaks higher than 12+. Group B (4
MCs) presents a unimodal distribution of $<$Q$>$Fe with its peak
higher than 12+. In groups C (29 MCs) and D (48 MCs), the
$<$Q$>$Fe remains higher and lower than 12+ throughout
\textit{ACE} passage through the MC, respectively. Possible
explanations to these distributions are discussed.

\end{abstract}

\keywords{Sun: coronal mass ejections (CMEs) $-$ magnetic
reconnection $-$ Sun: flares }

\section{INTRODUCTION}
Coronal mass ejections (CMEs) are the most energetic eruption in
the solar system. After CMEs propagate into interplanetary space,
they are called interplanetary coronal mass ejections (ICMEs),
which can cause geomagnetic activities and affect the satellites,
power grids and GPS navigation systems when they interact with the
geo-magnetosphere (Gosling et al. 1991; Zhang et al. 2003; Zhang
et al. 2007). ICMEs are extensively investigated by solar/space
physics community (Gopalswamy 2006; Jian et al. 2006; Kilpua et
al. 2013; Chi et al. 2015) for their important role in space
weather studies. Magnetic clouds (MCs, Burlaga et al. 1981), large
interplanetary magnetic flux rope structures embedded in ICMEs,
are more attractive as they usually cause stronger geomagnetic
storms compared to those ICMEs without MCs (Wu \& Lepping 2011,
2015; Lepping et al. 2011, 2015), and also because they have
relatively regular magnetic structures and can be fitted with flux
rope reconstruction methods (Lepping et al. 1990; Hu \& Sonnerup
2002; Wang et al. 2015).

By investigating the properties of MCs with in situ observations,
one can reveal the flux rope formation and eruption processes in
the corona. However, most MC parameters, e.g., the density,
temperature, velocity, volume and morphology, as well as magnetic
field strength, will experience large variation during the
propagation from near the Sun to 1 AU due to their expansion,
acceleration/deceleration, and interaction with other
interplanetary structures. This complicates efforts to investigate
the flux rope formation and eruption in the corona through in situ
data.

Fortunately, the charge states of ions inside MCs are frozen
before the flux ropes leave the corona (e.g., Rakowski et al.
2007; Ko et al. 2010; Lynch et al. 2011; Gruesbeck et al. 2011;
Lepri et al. 2012) and therefore may reflect their evolution
history near the Sun. Usually, the appearance of high ionic charge
states implies high electron temperature and is attributed to
magnetic reconnection occurring along the current sheet connecting
the flux rope to flare loops (Bemporad et al. 2006; Ko et al.
2013; Song et al. 2015b). As reconnection continues toward higher
altitude, the flux rope will be added with more layers consisting
of reconnected magnetic field lines. And the heated plasma will
correspondingly fill in the flux rope structure along these field
lines giving rise to an onion-layer morphology (Ko et al. 2013).
Therefore, high ionic charge states inside MCs may provide direct
signatures of reconnection in the post-CME current sheet, which is
very helpful to address some important unresolved issues of CMEs.

One heated debate associated with CME is that whether the flux
ropes exist prior to the eruption or they are formed during the
eruption (Song et al. 2014a, and references therein). Some
numerical simulations (Chen 1996; Lin \& Forbes 2000) and
observations (Zhang et al. 2012; Patsourakos et al. 2013; Cheng et
al. 2013, 2014; Song et al. 2014b, 2015b) support that flux ropes
can exist prior to the eruption, while other simulations
({Miki\'c} \& Linker 1994) and observations (Song et al. 2014a;
Ouyang et al. 2015) demonstrate that flux ropes can also be formed
during the eruption. When the eruption involves a pre-existing
flux rope with relatively low temperature, which is surrounded by
heated plasmas formed through magnetic reconnection in the current
sheet, we would expect to detect an MC with low-ionization-state
center and high-ionization-state shell (Lin et al. 2004; Ko et al.
2013; Song et al. 2015b). On the contrary, when the flux rope is
mostly formed during the eruption, a more ionized center might be
observed due to stronger heating and higher densities at lower
heights (Ko et al. 2013). It has been reported that the
[Fe$_{XVIII}$] line intensity in the post-CME current sheet
decreases with height, supporting this suggestion (Ciaravella et
al. 2013).

Iron is an astrophysically abundant heavy element. Its heavy mass
allows the iron ions, detected in situ, to be well separated from
other abundant heavy ions in the normal solar wind and ICMEs due
to significantly less interference from other species. This is
because ions are distinguished based on their mass and charge
(mass/charge) by present instruments (Gloeckler et al. 1998; von
Steiger et al. 2000). It has been suggested that Fe ions are more
affected by the continuing heating along a more extended space of
the plasma flow near the Sun, and other species such as C and O
ions are mainly sensitive to the heating processes at the earlier
stage (e.g., Ko et al. 2010; Gruesbeck et al. 2011; Lepri et al.
2012). Therefore, the in situ charge states of different ion
species can not be compared with each other in a straightforward
manner. Here we focus our study on Fe ions. Average Fe charge
states ($<$Q$>$Fe) in the solar wind are typically around 9+ to
11+ (e.g., Lepri et al. 2001). Based on the comparisons of charge
state distributions between the normal solar wind and ICMEs with
high Fe charge states, Lepri \& Zurbuchen (2004) determined a
reasonable threshold of $<$Q$>$ Fe as 12+, above which it implies
a considerable amount of high Fe charge states in ICMEs and vice
versa. High Fe charge states above this threshold have been
observed in over 50\% ICMEs, therefore it has been used as one
complementary identifier of ICMEs (Lepri et al. 2001; Lepri \&
Zurbuchen 2004). However, a comprehensive analysis of $<$Q$>$Fe
distribution inside MCs remains absent.

In this paper, we make a comprehensive survey of $<$Q$>$Fe
distributions of 96 MCs for solar cycle 23 using ACE data and
$\sim$96\% of them (92 MCs) can be classified into four groups,
shedding more light on the flux rope formation process. The
instruments are introduced in Section 2, and the data analysis is
presented in Section 3. Section 4 discusses possible explanations
to the results, which is followed by a summary in Section 5.

\section{INSTRUMENTS}

The Fe charge state is obtained using the Solar Wind Ion
Composition Spectrometer (SWICS, Gloeckler et al. 1998) on board
the \textit{Advanced Composition Explorer (ACE)} that launched in
1997 and orbits around the L1 point. As a mass spectrometer, SWICS
consists of a time-of-flight (TOF) system and an energy resolving
solid-state detector (SSD). The TOF system measures the speed of
each ion and the SSD determines its residual energy, which allow
the determination of mass, charge, and energy of the detected ion.
SWICS can determine the density, bulk speed, and the thermal speed
of almost 40 heavy ions and provide the charge state distributions
and abundances of $\sim$ 10 elements (also see Lepri et al. 2001).
In this study, we use the newly released SWICS 1.1 level 2 data,
which are calculated by an improved calibration algorithm and have
better accuracy from the previous versions (Shearer et al. 2014).
The process for removing accidental coincidence events has been
improved, leading to an increase in identification of rare ions
such as Fe$^{6+}$ and Fe$^{7+}$. Except the ionic charge state, we
also use the data from MAG (Smith et al. 1998) and SWEPAM (McComas
et al. 1998) on board \textit{ACE} for the magnetic field and
plasma measurements of associated MCs.

\section{DATA ANALYSIS}
MCs in the solar wind can be identified if the magnetic field
strength is higher than the surrounding background, the field
direction rotates smoothly through a large angle, and the proton
temperature is lower compared to the environment (Burlaga et al.
1981). In this study, we choose to use the MC lists published by
other authors, instead of examining MCs by ourselves. This
separates issues related to the MC identification and to the ionic
charge state analysis, making our study more focused. As no unique
and fully objective way has been developed to identify MCs in
interplanetary space (e.g., Huttunen et al. 2005), some subjective
judgements may influence the decision on whether a candidate is an
MC or not, as well as exact locations of its front and back
boundaries. Therefore, some disagreements on MC boundary locations
often exist among different MC lists. We will discuss how this may
affect our statistical results in Section 4.

\subsection{MC Lists}

The public 1-h resolution $<$Q$>$Fe data from SWICS are available
on the website of the \textit{ACE} science center since February
1998 ($http://www.srl.caltech.edu/ACE/ASC/index.html$), so we
include the MCs detected at L1 between February 1998 and December
2009 in this survey. Two MC lists are used here, one is published
by Huttunen et al. (2005) that included 2 MCs analyzed with
\textit{WIND} data in February 1998 at L1 point and 60 MCs
analyzed with \textit{ACE} data from March 1998 to December 2003
(See Table 2 of their paper). The other one is from \textit{WIND}
MFI team, which provided MCs from February 1995 to December 2009
($http://wind.nasa.gov/mfi/mag\_cloud\_S1.html$; Lepping \& Wu
2010), including 38 MCs from 2004 to 2009 observed at L1 point.
Note \textit{WIND} spacecraft didn't always orbit around L1 from
1994 to 2003, so we didn't use its MC information between 1998 and
2003. Totally, with the above two lists we acquired 100 MCs from
February 1998 to December 2009 at L1 point. Among them, 96 MCs
have complete Fe charge-state data as recorded by SWICS, and 2/2
MCs have incomplete/no Fe charge-state data. The typical scale of
MC structures is around 0.25 AU (e.g., Lepping et al. 2006),
$\sim$ 5900 Earth radii (R$_{E}$), and the average duration of MCs
is close to 20 h (Wu \& Lepping, 2011, 2015). When both
\textit{WIND} and \textit{ACE} are around L1 point, their
separation distance is mostly along the radial direction and
generally less than 240 R$_{E}$, so except for a time delay of
several tens of minutes, their plasma and magnetic field
measurements are almost identical. In some cases, there is a
lateral separation of about 200 R$_{E}$ along the east-west
direction, which is still significantly smaller than the typical
size of the MC near 1 AU, and not considerable. We compared the
\textit{WIND} and \textit{ACE} measurements for each MC from 2004
to 2009 and confirmed this point. Therefore, the magnetic field
measurement by \textit{WIND} and the charge-state measurement by
\textit{ACE} can be used together for the purpose of our study.

\subsection{Criteria and Statistical Results}
Following Lepri \& Zurbuchen (2004), the $<$Q$>$Fe higher/lower
than 12+ is defined as high/normal charge state. With this
threshold, we found that there are 50 MCs that contain high charge
state and 48 MCs not. This is consistent with the ICME percentage
associated with high Fe charge state as reported by Lepri et al.
(2001) and Lepri \& Zurbuchen (2004). In Figure 1, the histograms
display the yearly total number of MCs. The blue/red portions of
the bars correspond to MCs without/with high charge state. The
line connected by black filled circles shows the yearly sunspot
number. Figure 1 shows that the high Fe charge-state MCs present
in the rising (1998-1999), maximum (2000), and declining
(2001-2005) phases of solar cycle 23, and almost no high Fe
charge-state MCs are observed in the past solar minimum
(2006-2009).

After inspecting the $<$Q$>$Fe distributions inside the 96 MCs
with complete Fe charge-state data from SWICS, we find that
$\sim$96\% of them (92 MCs) can be divided into 4 groups (A, B, C,
and D) based on their distribution characteristics. The rest 4 MCs
are included in Group E, which will be discussed later. Table 1
lists the MC durations in every group according to their time
sequence. A summary of the classification results is presented in
the pie diagram of Figure 2.

To illustrate the $<$Q$>$Fe distribution characteristics inside
MCs for Groups A -- D, we select one representative event from
each group, and plot their corresponding observations of the
magnetic field, plasma, and Fe charge state in Figures 3-6. Panel
(a) presents the total magnetic field strength (black line) and
the three components, which are plotted in the Geocentric Solar
Ecliptic coordinate. The red, green, and blue lines correspond to
the X, Y, and Z components, respectively. The solar wind speed,
density, and temperature are presented in panels (b)-(d)
sequentially. Panel (e) is the Fe charge state distribution, and
the last panel is the $<$Q$>$Fe in which the horizontal purple
dotted line marks the level of 12+. The locations of the ICME
shock (if exists) and MC boundaries are depicted with purple
vertical solid line and dot-dashed lines, respectively.


\begin{table}[!htbp]
\begin{minipage}[t]{\columnwidth}
\caption{Classification groups (G) and their MCs at L1 point for
solar cycle 23.}
\renewcommand{\thefootnote}{\alph{footnote}}
\renewcommand{\footnoterule}{}
\tabcolsep=3pt
\begin{tabular}{llclcl}
  \hline
  G & & & MC start -- end time (UT) & &\\
    & & & yy/mm/dd -- mm/dd hh:mm & &\\

  \hline
  A & 98/10/19 04:00 -- 10/20 06:00 & & 99/04/16 20:00 -- 04/17 18:00& &99/04/21 12:00 -- 04/22 13:00  \\
    & 00/08/12 05:00 -- 08/13 02:00 & & 00/11/06 22:00 -- 11/07 15:00& &01/03/19 22:00 -- 03/21 23:00  \\
    & 01/05/28 11:00 -- 05/29 06:00 & & 01/11/24 17:00 -- 11/25 13:00& &02/04/17 24:00 -- 04/19 01:00  \\
    & 03/08/18 06:00 -- 08/19 11:00 & & 04/11/09 20:54 -- 11/10 03:24& &  \\

  B & 01/03/04 16:00 -- 03/05 02:00 & & 03/03/20 13:00 -- 03/20 22:00 & & 04/08/29 18:42 -- 08/30 20:48 \\
    & 05/12/31 14:48 -- 01/01 10:48 & &                               & &                               \\

  C & 98/06/24 12:00 -- 06/25 16:00\footnotemark[1]\footnotetext[1]{With one or a few data points lower than 12+.}  & & 98/09/25 08:00 -- 09/26 12:00\footnotemark[1] & & 99/02/18 14:00 -- 02/19 11:00 \\
    & 00/02/12 12:00 -- 02/12 24:00 & & 00/02/21 14:00 -- 02/22 12:00\footnotemark[1] & & 00/07/11 23:00 -- 07/13 02:00 \\
    & 00/07/15 05:00 -- 07/15 14:00 & & 00/07/15 19:00 -- 07/16 12:00 & & 00/09/17 23:00 -- 09/18 14:00 \\
    & 00/10/13 17:00 -- 10/14 13:00 & & 01/03/27 22:00 -- 03/28 05:00 & & 01/04/12 10:00 -- 04/13 06:00 \\
    & 01/04/28 24:00 -- 04/29 13:00 & & 01/10/03 01:00 -- 10/03 16:00 & & 02/02/28 18:00 -- 03/01 10:00 \\
    & 02/03/19 22:00 -- 03/20 10:00\footnotemark[1] & & 02/03/24 10:00 -- 03/25 12:00\footnotemark[1] & & 02/04/20 13:00 -- 04/21 15:00\footnotemark[1]\\
    & 03/10/29 12:00 -- 10/30 01:00\footnotemark[1] & & 03/11/20 11:00 -- 11/21 01:00 & & 04/04/04 02:48 -- 04/05 14:48 \\
    & 04/07/22 15:24 -- 07/22 23:06\footnotemark[1] & & 04/07/24 12:48 -- 07/25 13:18 & & 04/11/08 03:24 -- 11/08 16:36\footnotemark[1] \\
    & 04/11/10 03:24 -- 11/10 11:06 & & 05/05/15 05:42 -- 05/15 22:18 & & 05/05/20 07:18 -- 05/21 05:18\footnotemark[1] \\
    & 05/06/12 15:42 -- 06/13 07:06\footnotemark[1] & & 06/12/14 22:48 -- 12/15 19:48 & &                               \\

  D & 98/02/04 05:00 -- 02/05 14:00 & & 98/02/17 10:00 -- 02/18 04:00 & & 98/03/04 15:00 -- 03/05 21:00 \\
    & 98/06/02 10:00 -- 06/02 16:00 & & 98/08/20 08:00 -- 08/21 18:00\footnotemark[2] & & 98/11/13 04:00 -- 11/14 06:00 \\
    & 99/03/25 16:00 -- 03/25 23:00 & & 99/08/09 10:00 -- 08/10 14:00 & & 99/08/22 12:00 -- 08/23 06:00 \\
    & 99/09/21 20:00 -- 09/22 11:00 & & 99/11/14 01:00 -- 11/14 09:00 & & 00/07/13 15:00 -- 07/13 24:00 \\
    & 00/07/31 22:00 -- 08/01 12:00 & & 00/08/10 20:00 -- 08/11 08:00 & & 00/10/03 15:00 -- 10/04 14:00\footnotemark[2]\footnotetext[2]{With one or a few data points higher than 12+.} \\
    & 01/04/21 23:00 -- 04/22 24:00 & & 01/06/18 23:00 -- 06/19 14:00 & & 01/07/10 17:00 -- 07/11 23:00 \\
    & 01/10/31 22:00 -- 11/01 18:00 & & 02/05/19 04:00 -- 05/19 22:00 & & 02/08/02 06:00 -- 08/02 22:00 \\
    & 02/09/30 23:00 -- 10/01 15:00 & & 03/01/27 01:00 -- 01/27 15:00\footnotemark[2] & & 05/07/17 15:18 -- 07/18 03:48 \\
    & 05/10/31 02:54 -- 10/31 20:24 & & 06/02/05 19:06 -- 02/06 13:06 & & 06/04/13 14:48 -- 04/13 20:48\footnotemark[2] \\
    & 06/04/13 20:36 -- 04/14 09:54 & & 06/08/30 21:06 -- 08/31 14:54 & & 06/09/30 08:36 -- 09/30 21:36 \\
    & 07/01/14 14:06 -- 01/15 06:54 & & 07/03/24 03:06 -- 03/24 16:54 & & 07/05/21 22:54 -- 05/22 13:36 \\
    & 07/11/19 23:24 -- 11/20 12:54 & & 07/12/25 15:42 -- 12/26 06:48 & & 08/12/17 03:06 -- 12/17 14:24 \\
    & 09/01/02 06:06 -- 01/02 15:06\footnotemark[2] & & 09/02/04 00:06 -- 02/04 10:54 & & 09/03/12 00:42 -- 03/13 00:42 \\
    & 09/06/27 15:18 -- 06/28 18:18 & & 09/07/21 03:54 -- 07/21 17:06 & & 09/09/10 10:24 -- 09/10 16:24 \\
    & 09/09/30 07:54 -- 09/30 16:54 & & 09/10/12 12:06 -- 10/12 16:48 & & 09/10/17 22:06 -- 10/18 07:24 \\
    & 09/10/29 05:12 -- 10/29 22:48 & & 09/11/01 08:48 -- 11/02 07:48 & & 09/12/12 19:48 -- 12/14 05:18 \\

  E & 98/05/02 12:00 -- 05/03 17:00 & & 98/11/08 23:00 -- 11/10 01:00 & & 00/10/28 24:00 -- 10/29 23:00 \\
    & 05/06/15 05:48 -- 06/16 07:48 & &                               & &                                \\

  \hline
\end{tabular}
\end{minipage}
\end{table}

In Group A (11 MCs), the $<$Q$>$Fe inside MCs presents a bimodal
distribution with its two peaks higher than 12+, and less than 12+
between the peaks as shown in Figure 3(f). It is obvious that a
considerable amount of high Fe charge states ($\sim$16+) appear in
this event (Figure 3(e)) resulting in the elevated $<$Q$>$Fe.
Group B (4 MCs) is defined when the $<$Q$>$Fe inside MCs shows a
unimodal distribution with its peak higher than 12+ as shown in
Figure 4(f). Note the $<$Q$>$Fe on either side of the peak is
lower than 12+. In Group C (29 MCs), the $<$Q$>$Fe remains higher
than 12+ during \textit{ACE} passage through the MC as shown in
Figure 5(f). Figure 5(e) shows that this event is always
associated with lots of Fe charge states as high as $\geq$ 16+,
and the enhancement of high Fe charge states is almost coincident
with the passage of the MC. Note the MC $<$Q$>$Fe in Group C can
have a broad variation (e.g., 13+$\sim$16+) with different
patterns. In this study, we do not distinguish these pattern
details and classify them into one single group, since we are
mainly interested in the appearance of a considerable amount of Fe
ions with high charge states, which is defined when the $<$Q$>$Fe
is over 12+ (Lepri et al. 2001), and higher $<$Q$>$Fe values
simply indicating high charge state Fe ions are more abundant.
Group D (48 MCs) includes MCs not associated with high charge
states, i.e., with $<$Q$>$Fe less than 12+, as presented in Figure
6. Again, the MC $<$Q$>$Fe in Group D may have some variation as
well (e.g., 9+$\sim$11+), yet not of interest here.

Panels (e) in Figures 3-6 clearly show that when the $<$Q$>$Fe is
larger than 12+, a considerable amount of ions with charge states
even higher than 16+ are frequently present. Accordingly, we find
that the variation trend of $<$Q$>$Fe is in general consistent
with that of the Fe$^{\geq16+}$ abundance (i.e.,
Fe$^{\geq16+}/$Fe$_{total}$, not included here), which is often
used as a measure of the unusually enhanced charge states (e.g.,
Lepri et al. 2001). This supports that the prescribed threshold of
12+ is an appropriate criterion to represent the presence of a
considerable amount of high Fe charge states, which can be further
verified by checking the complete Fe charge state distributions
and $<$Q$>$Fe variations (cyan) within all MCs in Groups A-D as
shown in Figure 7. The horizontal white dotted line in each panel
marks the level of 12+ for $<$Q$>$Fe. Note only the MC part of the
data is plotted with the left/right boundary of each panel being
the MC start/end time. It is seen that the Fe charge state
distributions present an obvious bi-modal distribution, mainly
concentrating around 16+ and/or 10+. In Groups A and B, their
distributions appear around 16+ and 10+ at different intervals,
with undulating variation fluctuating with the $<$Q$>$Fe trend.
Note that if just one or a few discrete $<$Q$>$Fe data points are
lower/higher than 12+, i.e., their surroundings are higher/lower
than 12+, the discrete ones will be disregarded. In Group C/D,
most of the distributions are mainly characterized by charge-state
concentration around 16+/10+, consistent with their $<$Q$>$Fe
keeping higher/lower than 12+. This confirms that the 12+
condition is indeed an efficient identifier to separate the MC
events with high or normal Fe charge state (Lepri \& Zurbuchen,
2004). It is of some importance for relevant studies as the SWICS
on board \textit{ACE} suffered a hardware anomaly in 2011 and only
$<$Q$>$Fe data are available since then. Note one or a few
discrete data points in Group C/D are below/beyond 12+ (see notes
in Table 1 and corresponding panels in Figure 7), which are
disregarded as mentioned above. Two events (20020420 and 20050520)
in Group C have 5 continuous data points (less than 22\% of their
total data points) lower than 12+ near one boundary, while they do
not change the $<$Q$>$Fe distribution characteristics obviously
and are also neglected.

As mentioned, most MCs ($\sim$96\%) can be divided into Groups A
-- D. Four events can not be classified into these groups and are
included in Group E. We will discuss possible explanations for all
groups in next section.

\section{DISCUSSION}
In general, ICMEs can either exhibit high Fe charge states or not.
Lynch et al. (2011) derived the ionic charge state composition
distribution using axisymmetric MHD simulations of CMEs initiated
via either flux-cancellation or magnetic breakout mechanism. They
concluded that enhanced heavy ionic charge states within the flux
rope are a direct consequence of flare heating in the lower
corona, and not due to the heating of breakout reconnection. This
is reasonable as the breakout reconnection mainly removes the
constraint of overlying loops to trigger the eruption without
adding new layers to the flux rope (Antiochos et al. 1999). Lepri
and Zurbuchen (2004) suggested that magnetic connectivity of the
part of the ICME observed in situ to the flaring region is the key
to the presence of high Fe charge states. Here, we propose a
different scenario.

According to the CME model (Forbes \& Acton 1996; Lin \& Forbes
2000; Lin et al. 2004), the flux rope and the flare region can be
related with each other through the reconnecting current sheet in
the wake of a CME. The Fe ions with high/normal charge states can
be generated in the high/normal temperature current sheet (e.g.,
Ciaravella et al. 2013), in addition to the flare region, and fill
in the corresponding layers of the flux rope like `layers of an
onion' as mentioned (Ko et al. 2013). In this paper, the
high/normal temperature in the current sheet is defined as it is
beyond/below 2 MK. This is consistent with the early calculation
of Arnaud \& Raymond (1992), which showed that high Fe charge
states (${\geq15+}$) can have a considerable amount ($>30\%$)
above 2 MK. We point out that the temperature of a flux rope will
likely decrease with its expansion when propagating outward,
therefore, the $<$Q$>$Fe within the flux rope when reaching the
charge-state freezing distance may not be as high as that in the
earlier reconnecting current sheet region (e.g., Rakowski et al.
2007; Ko et al. 2010). Based on these points, we present our
understanding to the above observations in Figure 8 with some
schematics.

The left panels show the eruption with a pre-existing flux rope as
delineated with a yellow circle filled with blue, which means its
$<$Q$>$Fe is lower than 12+. The large purple circle depicts the
boundary of the MC. In panel (a1), the current sheet has a high
temperature, this leads the $<$Q$>$Fe to be higher than 12+, as
presented with red. As the ionic charge states are mostly frozen
before they leave the corona (e.g., Rakowski et al. 2007; Ko et
al. 2010), the measured MC will likely contain a low/high ionized
center/shell. If the spacecraft passes through the MC along the
upper green arrow, a bimodal distribution like Figure 3(f) will be
detected. If the spacecraft passes through the MC shell portion
along the lower arrow, a distribution shape as pointed with the
arrow should be observed, similar to Figure 5(f). Panel (a2)
describes the case that the current sheet temperature is high at
first, but becomes normal subsequently during the flux rope
formation. Then the MC will contain a high-ionization-state inner
shell and a low-ionization-state outer shell. The obtained
$<$Q$>$Fe shapes can be similar to Figures 3(f) or 4(f), depending
on the spacecraft passage. It is also possible to get a
distribution shape similar to Figure 6(f) if the spacecraft just
passes through the MC edge, which is not plotted in panel (a2).
Panel (a3) describes the case that the current sheet temperature
is normal during the flux rope growth. Then no Fe ions with high
charge states are available to fill in the flux rope, which
results in the distribution shape like Figure 6(f), no matter
where the spacecraft passes through the MC.

The right panels present the situation when the flux rope is
formed during the eruption, so just a large purple circle is
plotted at the MC boundary. Correspondingly, the top, middle, and
bottom panels refer to cases of which current sheet temperature
being high, first high then normal, and normal during the flux
rope formation. It is easy to understand that three $<$Q$>$Fe
distribution shapes can be obtained, except the bimodal
distribution as shown in Figure 3(f). Therefore, our analysis
suggests that the flux rope exists prior to the eruption if its MC
$<$Q$>$Fe distribution presents the bimodal shape. In summary, the
observed $<$Q$>$Fe distributions depend on how the spacecraft
crosses the MC as well as physical properties of the flux rope and
the current sheet near the Sun.

In order to support the above scenario, we checked the source
regions of 11 CMEs in Group A with EIT on board the \textit{Solar
and Heliospheric Observatory} and H$\alpha$ images from the Big
Bear Solar Observatory. The results show that 7 (Events 1-6 and 8)
of them are associated with filaments. This indicates a
pre-existing flux rope (Rust \& Kumar, 1994; Song et al. 2015a)
and supports our scenario. We can not identify the source region
for Event 7, and no filaments were observed in the source active
regions for Events 9-11. It is not easy to confirm or deny the
existence of other proxies of flux ropes prior to the eruptions,
such as sigmoid structures (Titov \& D\'emoulin, 1999) and/or hot
channels (Zhang et al. 2012; Song et al. 2015b) as no soft X-ray
and high temperature EUV observations available.

As mentioned, some disagreements on MC boundary locations often
exist among different MC lists (e.g., Huttunen et al. 2005). To
assess its influence to our statistical results, we checked
another MC list (http://wind.nasa.gov/index\_WI\_ICME\_list.htm)
on the \textit{WIND} website. As the \textit{WIND} locates at L1
point since 2004, we mainly compared the 31 MCs appeared both in
this list and our Table 1 between 2004 and 2009. The difference of
their start/end time varies in a range of 0$\sim$5.5/0.5$\sim$16.3
h with the average value being 1.7/4.5 h, while the average of MC
duration across the spacecraft is close to 20 h (e.g., Wu \&
Lepping, 2011, 2015). According to this MC list, only one event
(20050612) of the 31 MCs should be reclassified and moved from
Group C to B. This almost does not affect our statistics.

Considering the complex dynamical processes of magnetic
reconnection and CME eruption, as well as the propagation of flux
rope in interplanetary space, not all $<$Q$>$Fe distributions are
regular inside MCs and can be classified into Groups A -- D. Four
such events (Group E) are found, indicating some complicated cases
may exist. For example, when cold filament materials contained in
MCs were detected in situ (Gloeckler et al. 1999; Lepri et al.
2010), the $<$Q$>$Fe will descend obviously and the charge-state
distribution will change accordingly. The first event in Group E
is probably such a case (Gloeckler et al. 1999), in which the
$<$Q$>$Fe descends gradually from 16+ to 6+. The second event is
odd with former/later part lower/higher than 12+ if the two
discrete points beyond 12+ were neglected, and the rest two events
show three peaks higher than 12+ inside the MCs. These events can
not be explained with our simple and qualitative scenario.

We also note some high Fe charge states exist outside the MCs,
e.g., the sheath region in Figure 5(e). They might be generated by
processes such as the breakout reconnection or heating of a lower
coronal shock. They may also correspond to the outermost part of
the magnetic flux rope in the corona, but not regarded as parts of
MCs possibly because of their irregular magnetic structure as
measured in situ. For the high Fe charge states in ICMEs without
an embedding MC, we speculate that they may also be produced by
magnetic reconnection along the post-CME current sheet, yet the
ejecta does not evolve into a regular MC structure.

\section{SUMMARY}
We made a comprehensive survey of $<$Q$>$Fe distributions inside
96 MCs for solar cycle 23 using \textit{ACE} data. The high Fe
charge state is defined when the $<$Q$>$Fe is larger than 12+,
which means a considerable amount of Fe ions with high charge
states (e.g., $\geq$ 16+) appears inside the MCs. The statistical
results show that the distribution of 92 MCs ($\sim$96\%) can be
classified into four groups with different characteristics. Group
A (11 MCs) is defined when the $<$Q$>$Fe shows a bimodal
distribution with its double peaks higher than 12+. Group B (4
MCs) presents a unimodal distribution with its peak higher than
12+. Groups C (29 MCs) and D (48 MCs) represent that the $<$Q$>$Fe
keeps always higher and lower than 12+ during \textit{ACE} passage
through the MC, respectively.

A qualitative scenario was proposed to explain the results, which
can be used to infer the magnetic flux rope formation time and the
current sheet temperature information during eruption. The
high/normal Fe charge state indicates the current sheet
temperature is high/normal during the eruption, and the bimodal
distribution exists only if the flux rope has been formed prior to
the eruption. Our study supports this scenario in a preliminary
and statistical way. More detailed case studies are needed to
further test this scenario.

\acknowledgments We are grateful to the referee for his/her
constructive comments and suggestions, which improved the paper
greatly. We thank Susan T. Lepri, Lan Jian, Jun Lin, Yuming Wang,
Pengfei Chen, Yong Liu, and Chenglong Shen for their valuable
comments and discussion. We acknowledge the use of data from the
\textit{ACE} and \textit{WIND} missions. This work is supported by
the 973 program 2012CB825601, NNSFC grants 41274177, 41274175, and
41331068. J.Z. is supported by US NSF AGS-1249270 and NSF
AGS-1156120. G. L. is supported by ATM-0847719 and AGS-1135432.
The work of L.Z. is supported by NSF grant AGS-1432100 and
AGS-1344835.

\clearpage

\begin{figure}
\epsscale{1.0} \plotone{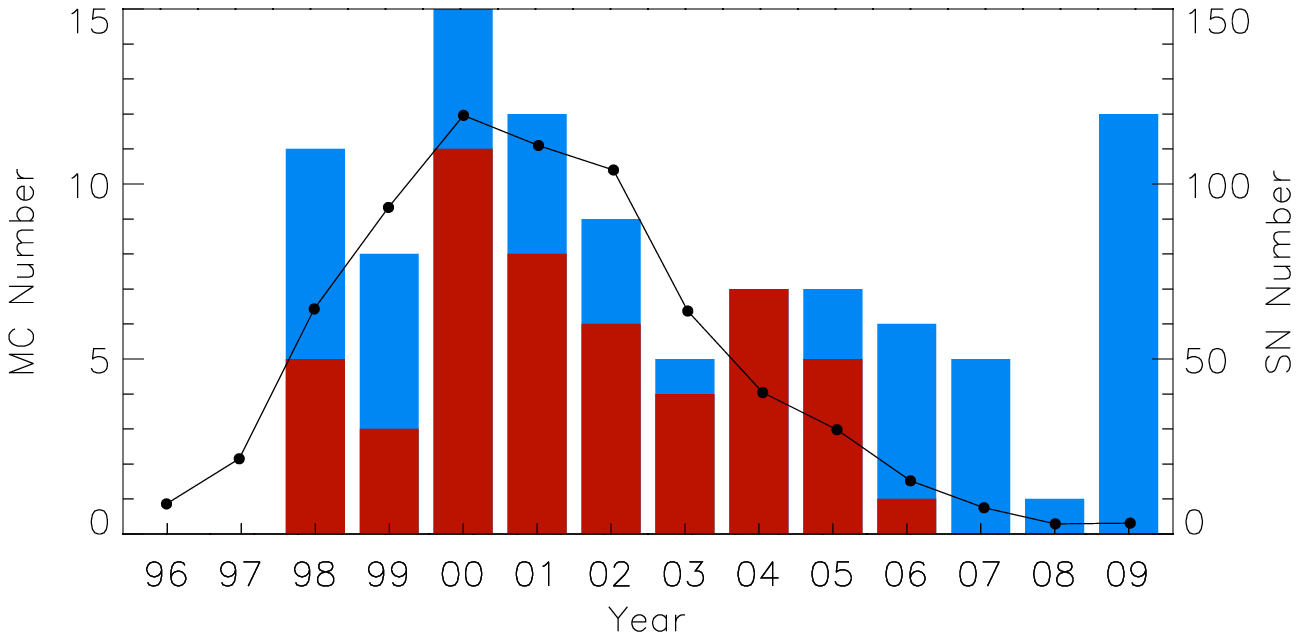} \caption{Yearly number of MCs in
the study. The blue/red portions of the bars correspond to the MCs
without/with high Fe charge state. The line connected by the
filled circles shows the yearly sunspot number. \label{Figure 1}}
\end{figure}

\begin{figure}
\epsscale{1.0} \plotone{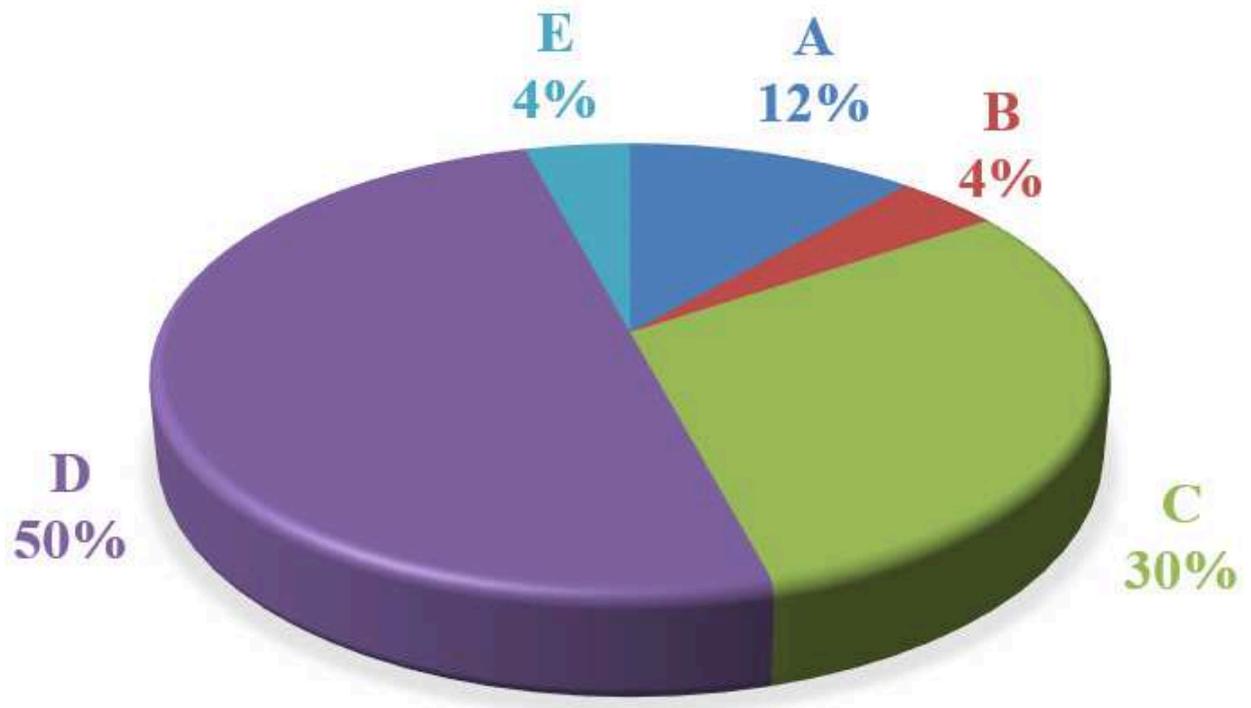} \caption{Pie diagram that shows
the percentages of different groups for the 96 MCs with complete
Fe charge-state data. \label{Figure 2}}
\end{figure}

\begin{figure}
\epsscale{0.85} \plotone{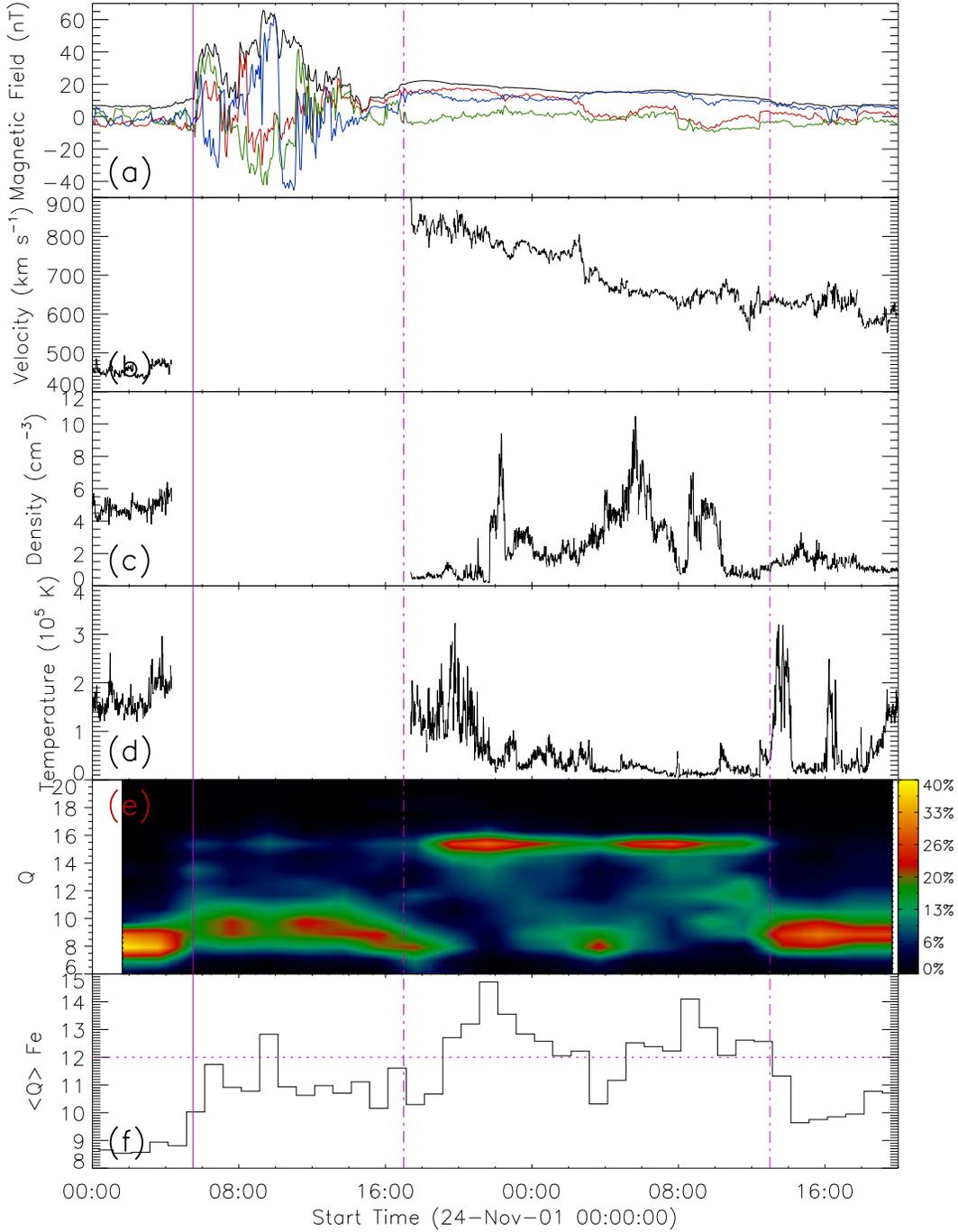} \caption{Solar wind parameters
and charge-state distribution measured with \textit{ACE} for
20011124 event in Group A. (a) the total magnetic field strength
(black line) and the X (red), Y (green), and Z (blue) components
in the GSE coordinate, (b)-(d) the bulk speed, density, and
temperature of solar wind, (e) 3D Fe charge-state-distribution
map, and (f) the $<$Q$>$Fe. The horizontal purple dotted line
marks the level of 12+. The locations of the ICME shock and MC
boundaries are depicted with purple vertical solid line and
dot-dashed lines, respectively. \label{Figure 3}}
\end{figure}

\begin{figure}
\epsscale{0.85} \plotone{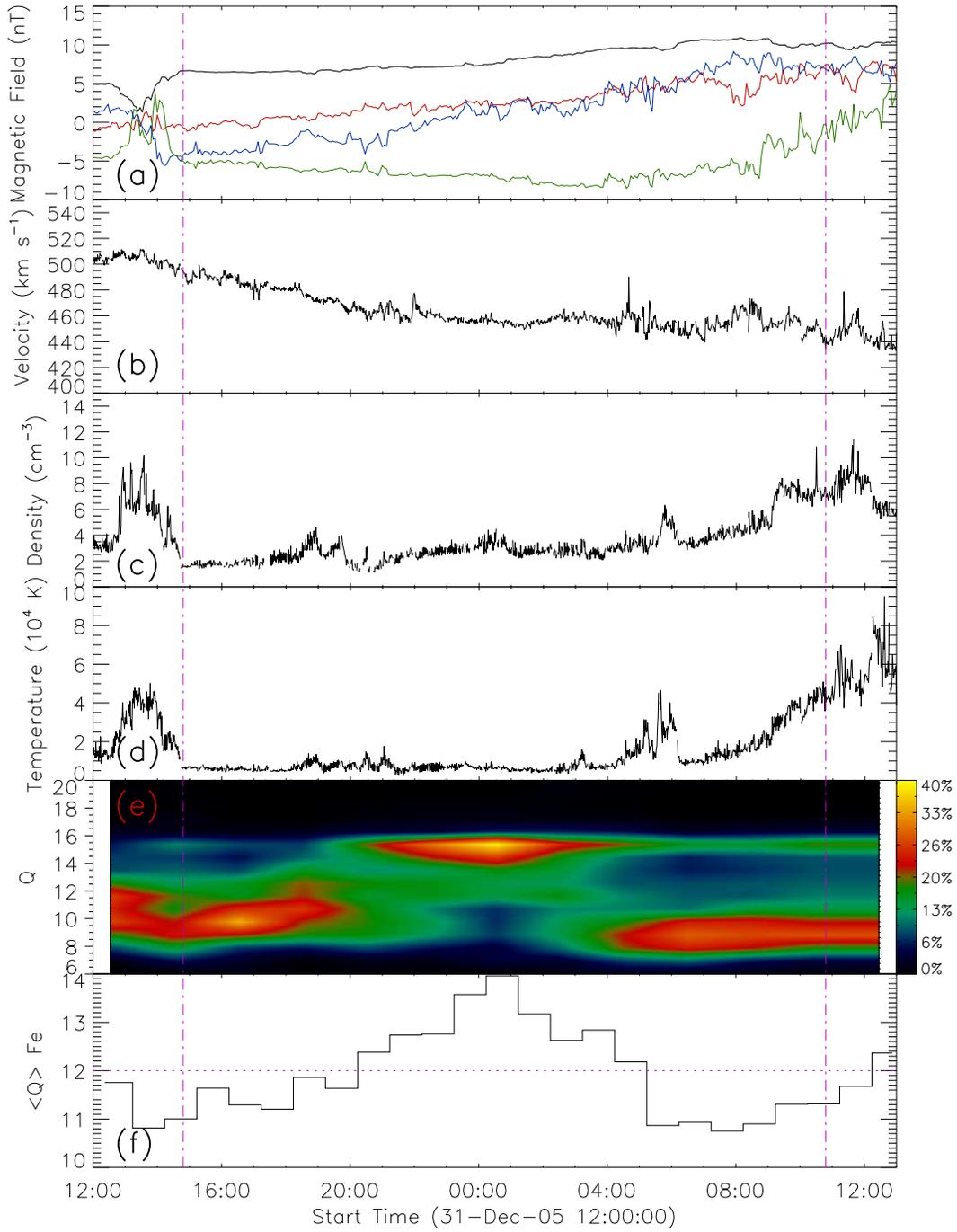} \caption{Same as Figure 3 but
for 20051231 event in Group B. \label{Figure 4}}
\end{figure}

\begin{figure}
\epsscale{0.85} \plotone{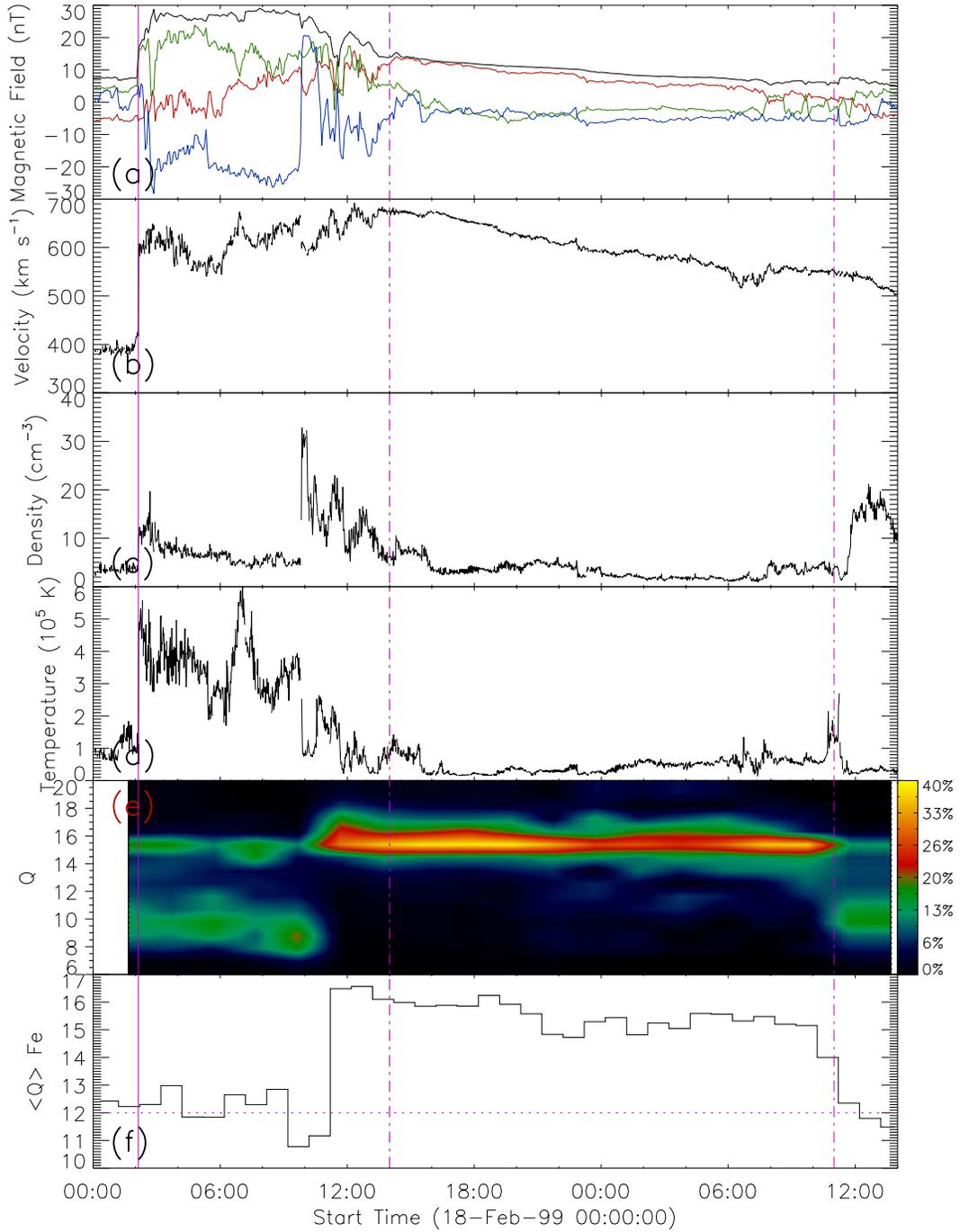} \caption{Same as Figure 3 but
for 19990218 event in Group C.  \label{Figure 5}}
\end{figure}

\begin{figure}
\epsscale{0.85} \plotone{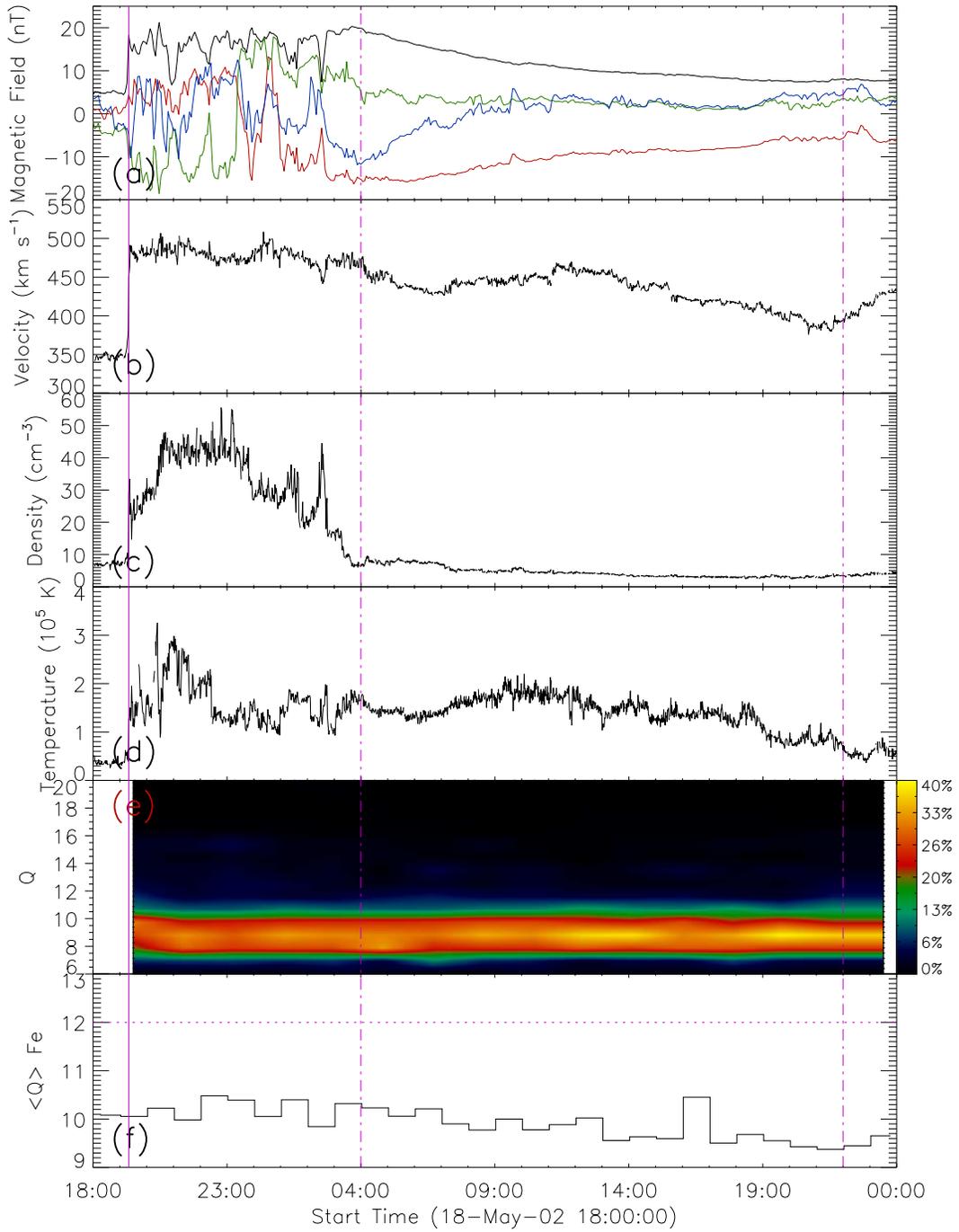} \caption{Same as Figure 3 but
for 20020519 event in Group D.  \label{Figure 6}}
\end{figure}

\begin{figure}
\epsscale{0.85} \plotone{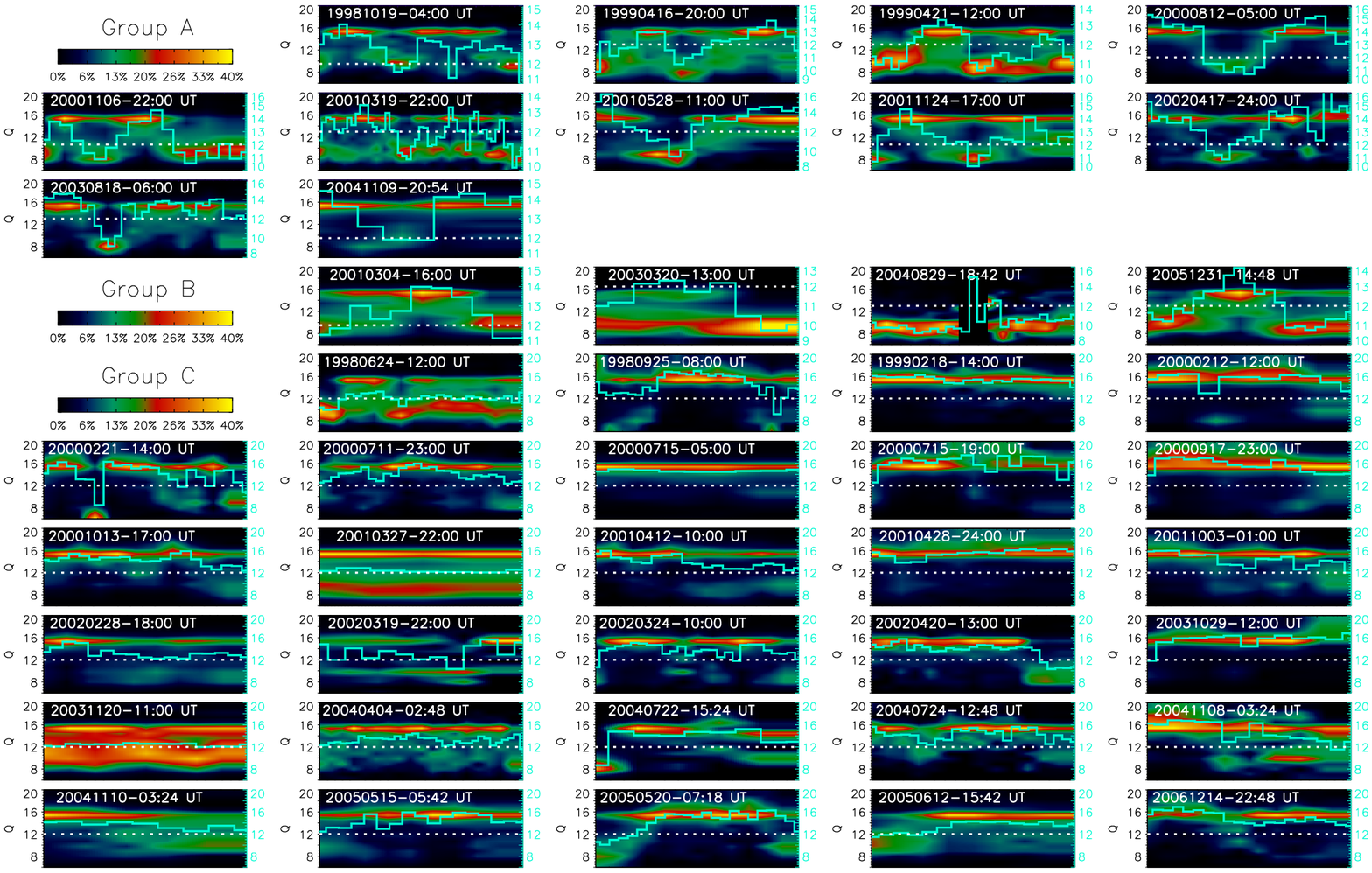}
               \plotone{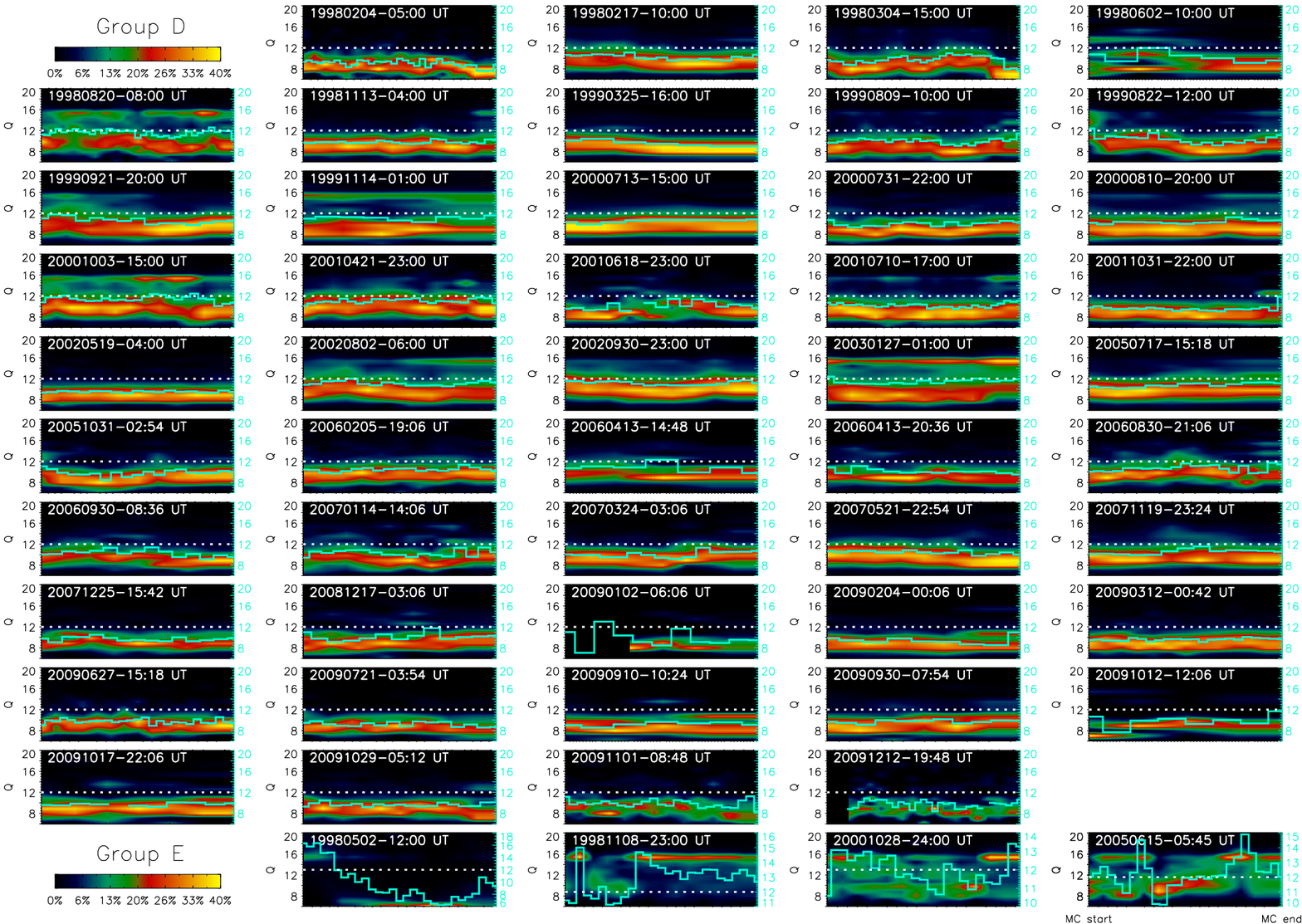}
               \caption{The Fe
charge state distributions and $<$Q$>$Fe variations (cyan) within
96 MCs. The horizontal white dotted lines mark the level of 12+
for $<$Q$>$Fe. Groups A, B, C and E are all associated with lots
of high Fe charge states, while Group D is mainly associated with
normal Fe charge states with $<$Q$>$Fe below 12+. The left/right
boundary corresponds to each MC start/end time. See text for
details. \label{Figure 7}}
\end{figure}

\begin{figure}
\epsscale{0.85} \plotone{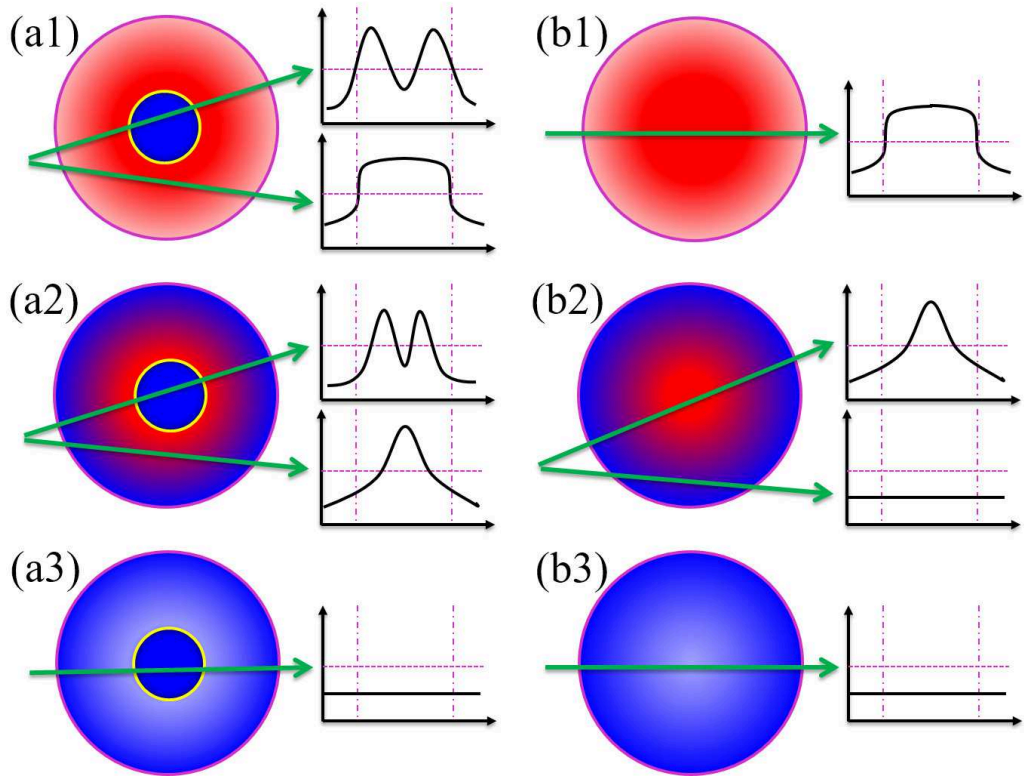} \caption{Schematic drawings of
the explanation for the $<$Q$>$Fe distributions inside MCs.
Red/blue denotes the $<$Q$>$Fe higher/lower than 12+. The
horizontal purple dotted lines in the coordinates mark the
$<$Q$>$Fe level of 12+, and the purple vertical dot-dashed lines
demarcate the MC boundaries. See text for details. \label{Figure
8}}
\end{figure}

\end{document}